\documentstyle[12pt]{article}
\begin{document}
\def\thefootnote{\fnsymbol{footnote}}
\begin{flushright}
KANAZAWA-05-16  \\
November, 2005
\end{flushright}
\vspace*{2cm}
\begin{center}
{\LARGE\bf Singlino dominated LSP as CDM 
candidate in supersymmetric models with an extra U(1)}\\
\vspace{1 cm}
{\Large Daijiro Suematsu}
\footnote[2]{e-mail:~suematsu@hep.s.kanazawa-u.ac.jp}
\vspace {1cm}\\
{\it Institute for Theoretical Physics, Kanazawa University,\\
        Kanazawa 920-1192, Japan}\\
\end{center}
\vspace{1cm}
{\Large\bf Abstract}\\
We consider a singlino dominated neutralino in supersymmetric
models with an extra U(1). In case both the $\mu$ term and also the $Z^\prime$ 
mass are generated by the vacuum expectation 
value of the scalar component of the same singlet chiral superfield, 
generically the lightest neutralino is not expected to be dominated 
by the singlino. 
However, if the gaugino corresponding to the extra U(1) is sufficiently 
heavy, the lightest neutralino can be dominated by the singlino and still 
satisfy the constraints resulting from the $Z^\prime$ phenomenology. 
We assume a supersymmetry breaking scenario in which the extra U(1) 
gaugino can be much heavier than other gauginos. 
In that framework we show that the singlino dominated lightest
neutralino may be a good candidate for dark matter in a parameter 
space where various phenomenological constraints are satisfied. 

\newpage
\setcounter{footnote}{0}
\def\thefootnote{\arabic{footnote}}
\section{Introduction}
Various astrophysical observations seem to confirm the existence of 
a substantial amount of non-relativistic and non-baryonic dark matter 
\cite{cmb,sdss}.
The amount of cold dark matter (CDM) has been estimated 
to be $\Omega_{\rm CDM}h^2=0.12\pm 0.01$ through combined 
analyses of the Solan
Digital Sky Survey (SDSS) data on the large scale structure and the
Wilkinson Microwave Anisotropy Probe (WMAP) data.
This fact suggests that the Standard Model (SM) of Elementary Particle
Physics is required to be
extended to include a CDM candidate.

Supersymmetric extensions of the SM have been considered to be the most 
promising candidate for a solution to the gauge hierarchy problem \cite{susy}. 
It is interesting that supersymmetric models can naturally contain a CDM
 candidate. If the $R$ parity is conserved, the lightest neutral
supersymmetric particle (LSP) is stable and then can be a good 
candidate for the CDM. The most
promising particle to play such a role is the lightest neutralino. 
Relic density of the thermally produced lightest neutralino 
is determined by its density 
at freeze-out temperature $T_F$. It is estimated by 
$H(T_F)\simeq \langle \sigma_{\rm ann}v\rangle n_\chi(T_F)$, where
$H(T_F)$ is the Hubble parameter at $T_F$ and
$\langle\sigma_{\rm ann}v\rangle$ is thermal average of 
annihilation cross section times relative velocity of
neutralinos \cite{xf}. Since neutralino $\chi$ has mass of the
order of the weak scale and feels only the weak interaction, 
we can generally expect its energy density $\Omega_\chi$ to be $O(1)$.
Detailed analyses of this relevant quantity have been extensively
done, especially, in the minimal supersymmetric standard model 
(MSSM) \cite{ann}. 
In both frameworks of the minimal supergravity and the constrained MSSM, 
many works have shown that the relic neutralino abundance can 
accommodate the observed $\Omega_{\rm CDM}h^2$ 
as long as model parameters are suitably selected \cite{ann1,mssm,mssm1}.

Although the MSSM is today the best candidate to describe physics 
beyond the SM, it has certain weak points too. Among them is 
the well known $\mu$ problem. 
The next to the MSSM (NMSSM) \cite{nmssmo} 
and the models with an extra U(1) \cite{extra,extra1} have been proposed as
elegant solutions of the $\mu$ problem. 
In both cases an SM singlet chiral superfield $\hat S$ 
is introduced and superpotential
is extended by a term $\lambda\hat S\hat H_1\hat H_2$.\footnote{In this 
paper we put a hat on the character describing a superfield. 
For its component fields, we put a tilde on the same 
character for superpartners of the SM fields 
and use just the same character without the hat for the SM fields.
Otherwise, the field without a tilde should be understood as a scalar
component.}
The $\mu$ term $\mu \hat H_1\hat H_2$ is generated as 
$\mu=\lambda\langle S\rangle$, {\it i.e.} by a vacuum expectation
value of the scalar component of $\hat S$ through the introduced operator.
A difference between the two models appears in the way a bare term 
$\mu \hat H_1\hat H_2$ is forbidden and the potential for $S$ is stabilized.
Related to these issues, a cubic term $\kappa \hat S^3$ and an extra
U(1) have been introduced in each case, respectively.
It is worth noting that there is a cosmological domain wall problem 
in the former case, which can be escaped by introducing suitable 
non-renormalizable operators \cite{nonre}. However, the models with an 
extra U(1) do not have such a problem.
Moreover the models with an extra U(1) often appear as the effective theory of
superstring. They generally contain exotic fields \cite{string}.
Since the Higgs and the neutralino sector in both the NMSSM 
and the models with an extra U(1) are extensions of those of the MSSM, 
the relic density of the lightest neutralino is
expected to show different features from the corresponding one in the MSSM.
Since these models have various interesting new aspects as the extensions 
of the MSSM, it is worth studying them in detail on the basis of the
WMAP data.

In ref.~\cite{nmssm0} the relic density of the lightest neutralino 
in the NMSSM has been studied and its compatibility with the WMAP results has
been discussed in detail. 
Since the Higgs sector is extended by scalar and pseudoscalar
components of the singlet chiral superfield $\hat S$ as compared 
with the MSSM, annihilation of the lightest
neutralino can have various effective modes.
As the result, the WMAP constraint can be satisfied easier in the NMSSM 
than in the MSSM. This happens in various cases such that the lightest 
neutralino is the bino-like LSP, the bino-Higgsino mixed LSP 
and the singlino-like LSP. In particular, the singlino-like LSP is shown to
annihilate effectively by the effects of additional couplings $\lambda \hat
S\hat H_1\hat H_2$ and $\kappa\hat S^3$. The singlino LSP is found to 
satisfy the constraint from the relic density 
although it has no SM interactions.
The neutralino relic density has also been examined in the modified NMSSM
(nMSSM) where the cubic term $\kappa\hat S^3$ is replaced by a linear
term of $\hat S$ \cite{nmssm}.

The relic density of the lightest neutralino in a model with an extra U(1) 
has already been studied in \cite{ce}.
Since the neutralino sector is extended as compared with the MSSM 
by a fermionic component $\tilde S$ and also an extra U(1) 
gaugino $\tilde\lambda_x$, the features of the neutralinos can 
be different from the corresponding ones in both the MSSM 
and the NMSSM \cite{ce,sneut}. In particular, 
if the singlino $\tilde S$ dominates the lightest neutralino, 
a large change is expected to appear 
in the neutralino phenomenology. The relic density
needs to be studied by taking account of such a situation.
In the simple models with an extra U(1), the extra U(1) symmetry is 
supposed to be broken by a vacuum expectation value $\langle S\rangle$, which 
gives the origin of the $\mu$ term.
The lightest neutralino dominated by the singlino component is expected 
to occur when $\langle S\rangle$ takes a value of the order of the 
weak scale as long as $\lambda$ is not so small.\footnote{
In the NMSSM the singlino domination of the LSP for a larger value of 
$\langle S\rangle$ is also studied in \cite{nmssm0}. There it is shown that 
even in that case the singlino-like LSP is possible for a 
very small $\lambda$.} 
In the NMSSM such a value of $\langle S\rangle$ brings no problem and
the singlino dominated LSP can realize the CDM abundance as discussed
in \cite{nmssm0}.
In the models with an extra U(1), however, there exist severe constraints 
on $\langle S\rangle$ resulting from the mass of the extra U(1) gauge boson 
$Z^\prime$ and its mixing with the ordinary $Z$ boson
based on the direct search and the electroweak precision
measurements \cite{pdg,kmix1}.
These constraints tend to require that $\langle S\rangle$ should be
more than $O(1)$~TeV as long as we do not consider a special
situation.\footnote{Even in the extra U(1) models, if one considers 
a model with a secluded singlet sector, which is called the $S$-model 
in \cite{secl}, $\langle S\rangle$ can take a value of the weak scale. 
In this case phenomenological features at the weak scale are very 
similar to the NMSSM with a weak scale $\langle S\rangle$.} 
Thus, we cannot expect a substantial difference in the lightest
neutralino sector from the MSSM since both $\tilde S$
and $\tilde\lambda_x$ practically decouple from the lightest neutralino. 
Here it is useful to note that the lightest neutralino can be 
dominated by the singlino even in this kind of models with an 
extra U(1) if the gaugino $\tilde\lambda_x$ can be very heavy. 
In that case we may have the lightest neutralino as a candidate for the CDM,
which has a very different nature from that in both the MSSM and the NMSSM.
From this point of view, the relic density of the lightest neutralino 
has been studied in \cite{ce}. 

In the models with an extra U(1) the singlino dominated
lightest neutralino feels the extra U(1) gauge interaction. Thus, it can
annihilate through the $s$-channel exchange of $Z^\prime$ even if it is
dominated by the singlino. 
Unfortunately, the result in \cite{ce} seems to show that the WMAP
constraint cannot be satisfied by the singlino-like LSP 
if we impose the currently known lower bound for the $Z^\prime$ mass.
However, their analysis has been done for the case that the lightest
neutralino is composed of 81\% singlino and 12.5\% Higgsinos
$\tilde H_{1,2}$ as a typical example.
If we assume that the extra U(1) gaugino $\tilde\lambda_x$ 
can be much heavier and
the lightest neutralino is almost dominated by the singlino, its
annihilation is expected to be enhanced since the extra U(1) charge of
the singlino $\tilde S$ can be generally larger than the Higgsinos 
$\tilde H_{1,2}$, as will be discussed later. 
From this viewpoint, it seems worth reanalyzing
this possibility by assuming much larger mass for $\tilde\lambda_x$
than the one assumed in \cite{ce}.
The lightest neutralino may have very small mass which is forbidden 
in the MSSM already. However, if it is dominated by the singlino, 
it can be expected to escape the current experimental constraints and 
be a good CDM candidate. Since the models with an extra U(1) are 
the interesting extension of the MSSM, it will be useful to 
reexamine the inherent possibility in such models. 

This paper is organized as follows.  
In section 2 we briefly review the features of the models with an extra
U(1) and discuss their neutralino sector in the case of a heavy extra 
U(1) gaugino $\tilde\lambda_x$. 
In section 3 we study numerically the features of the lightest neutralino 
and also estimate the relic abundance of the lightest neutralino in that case.
Then we examine the compatibility with the WMAP constraints. 
Section 4 is devoted to the summary. 
In the Appendix we present an example for the supersymmetry breaking 
scenario which can realize the assumed possibility of non-universal 
mass only for the Abelian gaugino. 
 
\section{Models with large Abelian gaugino mass}
We consider the models with an extra U(1), which contain 
a very heavy extra U(1) gaugino and can give a solution to 
the $\mu$ problem. 
We assume that the extra U(1) gauge symmetry is broken by the vacuum
expectation value $\langle S\rangle$ of the scalar component 
of the SM singlet chiral superfield $\hat S$. 
The $\mu$ term is considered to be generated through an operator in
the superpotential of the form
\begin{equation}
W_{\rm ob}=\lambda \hat S\hat H_1\hat H_2 +\dots,
\label{mu}
\end{equation}
where $\hat H_{1,2}$ are the ordinary doublet Higgs chiral superfields
and a coupling constant $\lambda$ is assumed to be real.
This superpotential requires that $\hat H_{1,2}$ also have 
extra U(1) charges $Q_{1,2}$, which satisfy a charge conservation
condition
\begin{equation}
Q_1+Q_2+Q_S=0.
\label{cons}
\end{equation}
As a result of this feature, if the scalar components of 
$\hat H_{1,2}$ and $\hat S$ obtain the vacuum expectation values 
defined by
\begin{equation}
\langle H_1\rangle=\left(\begin{array}{c}v_1 \\ 0\\ \end{array}\right),
\qquad
\langle H_2\rangle=\left(\begin{array}{c}0 \\ v_2\\\end{array}\right),
\qquad
\langle S\rangle =u,
\end{equation}
the neutral gauge bosons $Z_\mu$ and $Z^\prime_\mu$ mix with each other.
This mixing can be represented by a mass matrix $M_{ZZ^\prime}$
 as \cite{extra,kmix1}
\begin{equation}
\left(\begin{array}{cc}
{g_W^2+g_Y^2\over 2}v^2 & 
{g_x\sqrt{g_W^2+g_Y^2}\over 2}v^2(Q_1\cos^2\beta-Q_2\sin^2\beta) \\
{g_x\sqrt{g_W^2+g_Y^2}\over 2}v^2(Q_1\cos^2\beta-Q_2\sin^2\beta) &
{g_x^2\over 2}v^2(Q_1^2\cos^2\beta+Q_2^2\sin^2\beta+Q_S^2{u^2\over v^2}) \\
\end{array}\right)
\label{zmass}
\end{equation}
where we use the basis $(Z_\mu, Z_\mu^\prime)$ and
$v^2=v_1^2+v_2^2$ and $\tan\beta=v_2/v_1$.
An extra U(1) charge $Q_f$ and a coupling $g_x$ are 
defined through the covariant derivative
\begin{equation}
D_\mu=\partial_\mu+i{\tau^3\over 2}g_WW_\mu^3+i{Y_f\over 2}g_YB_\mu
+i{Q_f\over 2}g_xZ_\mu^\prime.
\label{covd}
\end{equation}
Mass eigenvalues of these neutral gauge bosons can be expressed as
\begin{eqnarray}
&&m_{Z_1}^2\simeq m_Z^2-m_Z^2{g_x\tan 2\xi\over\sqrt{g_W^2+g_Y^2}}
(Q_1\cos^2\beta-Q_2\sin^2\beta), \nonumber \\ 
&&m_{Z_2}^2\simeq
{g_x^2\over 2}(Q_1^2v_1^2+Q_2^2v_2^2+Q_S^2u^2)
+m_Z^2{g_x\tan 2\xi\over\sqrt{g_W^2+g_Y^2}}
(Q_1\cos^2\beta-Q_2\sin^2\beta), 
\label{z2}
\end{eqnarray}
where $m_Z$ is the $Z$ boson mass in the SM and 
$\xi$ is a $ZZ^\prime$ mixing angle.

Direct search for the new neutral gauge boson and precise
measurements of the electroweak interactions constrain the
mass eigenvalue $m_{Z_2}$ of the new gauge boson and the $ZZ^\prime$ mixing
angle $\xi$. These conditions may be summarized as 
$m_{Z_2}~{^>_\sim}~600$~GeV and $\xi~{^<_\sim}~10^{-3}$
\cite{pdg},\footnote{This bound for $m_{Z_2}$ obtained from the
$Z^\prime$ decay into the dilepton pairs depends on the models. 
It can be relaxed if $Z^\prime$
has a substantial decay width into non-SM fermion pairs such as 
neutralino pairs \cite{gkk}. This is expected to occur in the case 
that the singlino dominated neutralino is light enough to make this 
decay mode possible and has a larger coupling with $Z^\prime$ compared with 
the electrons. We will comment on this point later.}
which constrain the value of $u$ directly and also the value of $\lambda$ 
indirectly through the relation $\mu=\lambda u$.
The value of $\mu$ is restricted by the chargino mass bound 
and also the electroweak symmetry breaking condition \cite{extra,extra1}.
It is useful to note that the $ZZ^\prime$ mixing constraint disappears for a
special case $\tan\beta\simeq \sqrt{Q_1/Q_2}$. 
In this case we can regard the lower bound of $u$ as the one which comes
from the direct search of $Z^\prime$. 
Moreover, since $Q_1Q_2>0$ should be satisfied,
the charge conservation (\ref{cons}) for the extra U(1) makes $|Q_S|$
a larger value than other Higgsino charges $|Q_{1,2}|$. 
Since the interaction of the singlino with $Z^\prime$ can be larger compared 
with that of $\tilde H_{1,2}$, the annihilation of the singlino-like LSPs
through the $s$-channel exchange of $Z^\prime$ will be enhanced
as the singlino component in the LSP increases.
Although this situation may require tuning of supersymmetry 
breaking parameters, it may bring interesting
phenomenology different from that of the MSSM. 
Thus in the following study, we consider the situation approximated 
by this special condition as the first step and we 
only impose the constraint $m_{Z_2}\ge 600$~GeV. 

In this model the neutralino sector is extended into six components, 
since there are two additional neutral fermions 
$\tilde\lambda_x$ and $\tilde S$ compared with the MSSM.  
If we take the canonically normalized gaugino basis 
${\cal N}^T=(-i\tilde\lambda_x, -i\tilde\lambda_W^3, -i\tilde\lambda_Y, 
\tilde H_1, \tilde H_2, \tilde S)$ and define
the neutralino mass term as
${\cal L}_{\rm neutralino}^m=-{1\over 2}{\cal N}^T{\cal MN}+{\rm h.c.}$,
the 6 $\times$ 6 neutralino mass matrix ${\cal M}$ can be represented
as\footnote{Kinetic term mixing between two Abelian vector superfields 
is not considered here.
The study of their phenomenological effects can be found in \cite{sneut}.} 
\begin{equation}
\left( \begin{array}{cccccc}
M_x & 0 & 0 & {g_xQ_1\over \sqrt 2}v\cos\beta 
& {g_xQ_2\over \sqrt 2}v\sin\beta & {g_xQ_S\over \sqrt 2}u \\
0 & M_W & 0 &m_Zc_W\cos\beta & -m_Zc_W\sin\beta &0 \\
0 & 0 & M_Y & -m_Zs_W\cos\beta & m_Zs_W\sin\beta &0 \\
{g_xQ_1\over \sqrt 2}v\cos\beta &m_Zc_W\cos\beta &-m_Zs_W\cos\beta & 0 & 
\lambda u & \lambda v\sin\beta \\
{g_xQ_2\over \sqrt 2}v\sin\beta & -m_Zc_W\sin\beta & m_Zs_W\sin\beta 
& \lambda u & 0 & \lambda v\cos\beta \\
{g_xQ_S\over \sqrt 2}u & 0 & 0 & \lambda v\sin\beta & \lambda v\cos\beta & 0\\
\end{array} \right).
\end{equation}
Neutralino mass eigenstates $\tilde\chi_a^0(a=1\sim 6)$ are related 
to ${\cal N}_j$
by a mixing matrix $U$ as
\begin{equation}
\tilde\chi^0_a=\sum_{j=1}^6U_{aj}{\cal N}_j,
\label{meig}
\end{equation}
where $U$ is defined in such a way that 
$U{\cal M}U^T$ becomes diagonal.

Here we focus our attention to the composition of the lightest
neutralino. If $u$ can take a value similar to $v_{1,2}$ 
or less than those, the lightest neutralino is expected to be 
dominated by the singlino $\tilde S$ as in the case of the NMSSM
and the nMSSM \cite{nmssm0,nmssm}. The lightest neutralino with a sizable 
singlino component can be a good CDM candidate, if it can annihilate 
sufficiently well \cite{nmssm0,nmssm,ot}.
In the present model, however, the $Z^\prime$ constraints seem to 
require that $u$ is much larger than $v_{1,2}$ as mentioned before. 
As the result, $\tilde\lambda_x$ and $\tilde S$ tend to decouple from 
the lightest neutralino sector as long as the mass of $\tilde\lambda_x$ 
is assumed to be a similar value to other gaugino mass of $O(m_{1/2})$.
The composition of the lightest neutralino is similar to that of 
the MSSM. Then we cannot find distinctive features in the lightest 
neutralino sector in this case.

If there exists a large additional contribution to 
the gaugino mass $M_x$, for example, following a scenario 
discussed in the appendix, however, 
the situation is expected to change drastically. 
The lightest neutralino can be dominated by the singlino $\tilde S$.
In fact, if the gaugino $\tilde\lambda_x$ is heavy enough to satisfy 
$M_x \gg {g_xQ_S\over \sqrt 2}u$, we can integrate out 
$\tilde\lambda_x$ as in case of the seesaw mechanism. 
A resulting $5\times 5$ neutralino mass matrix can be expressed as
\begin{equation}
\left( \begin{array}{ccccc}
M_W & 0 &m_Zc_W\cos\beta & -m_Zc_W\sin\beta &0 \\
0 & M_Y & -m_Zs_W\cos\beta & m_Zs_W\sin\beta &0 \\
m_Zc_W\cos\beta &-m_Zs_W\cos\beta &
-{g_x^2Q_1^2\over 2M_x}v^2\cos^2\beta   & 
\lambda u & \lambda v\sin\beta \\
-m_Zc_W\sin\beta & m_Zs_W\sin\beta & \lambda u & 
-{g_x^2Q_2^2\over 2M_x}v^2\sin^2\beta & \lambda v\cos\beta \\
 0 & 0 & \lambda v\sin\beta & \lambda v\cos\beta & 
-{g_x^2Q_S^2\over 2M_x}u^2\\
\end{array} \right).
\label{mchi}
\end{equation}
This effective mass matrix suggests that the lightest neutralino 
tends to be dominated by the singlino $\tilde S$ 
as long as $M_{W,Y}$ and $\mu(\equiv\lambda u)$ are not smaller compared
with ${g_x^2Q_S^2u^2\over 2M_x}$. Since $M_W$ and $\mu$ cannot to be
less than $100$~GeV because of the lightest chargino mass bound, 
this condition is expected to be naturally satisfied in the case of 
$M_x\gg u$. 
In such a case, the phenomenology of the lightest neutralino can change
largely from that of the MSSM and also the NMSSM. 
We consider such a situation in the following.
  
\section{Singlino dominated neutralino dark matter}   
\subsection{Singlino dominated lightest neutralino}
In the present model, important parameters related to the neutralino 
sector are the gauge couplings $g_{W,Y}$, $g_x$, 
the gaugino mass $M_{W,Y}$, $M_x$,
the extra U(1) charges $Q_{1,2}$, $\tan\beta$, $u$ and the coupling $\lambda$,
which has the relation to $\mu$. 
We make several assumptions on these parameters to simplify
numerical analyses.
Firstly, we impose both the coupling unification and the gaugino mass 
universality for the MSSM contents.
Even if we impose the unification condition for the SM gauge couplings,  
there remains a freedom for normalization of the extra U(1) coupling
constant, which may be defined by $g_x=kg_Y$.
In the present analyses we fix it to be $k=1$. 
Secondly, we consider the case of $\tan\beta=\sqrt{Q_1/Q_2}$, which
automatically guarantees to satisfy the constraint from the 
$ZZ^\prime$ mixing. 
For the extra U(1) charge, we assume $Q_1=-4$ and $Q_2=-1$. This means
$\tan\beta=2$ and also the singlino can have a rather large 
charge $Q_S=5$.\footnote{The extra U(1) charge is normalized 
with a factor $1/2$ as
shown in (\ref{covd}). Under this normalization the charges of 
$\hat H_{1,2}$ used in \cite{ce} are $Q_{1,2}=2$.
The extra U(1) charge assignment is constrained by the anomaly 
free conditions. However, we do not go further into this issue here 
and we only assume $Q(f_L)=1$ for the left-handed quarks and 
leptons as a toy model, for simplicity.} 
Under these assumptions, there remain four free parameters 
$M_W$, $M_x$, $u$ and $\lambda$. 
We practice numerical analyses by varying these parameters.

We present the results of the numerical
analysis obtained by scanning these parameters in the region such as
\begin{equation}
300~{\rm GeV}\le u \le 2300~{\rm GeV}, \quad
3~{\rm TeV}\le M_x \le 115~{\rm TeV}, \quad 
0\le \lambda\equiv{\mu\over u}\le 0.75.  
\label{para}
\end{equation}
The last condition comes from the perturbative bound of the coupling 
$\lambda$ \cite{nmssm,higgsb}.
Throughout these analyses  we impose the constraints on the mass 
of the chargino \cite{chargino}, the extra gauge boson and 
the lightest neutral Higgs scalar:
\begin{equation}
m_{\chi^\pm}\ge 104~{\rm GeV}, \quad m_{Z_2}\ge 600~{\rm GeV}, \quad
m_h \ge 114~{\rm GeV}.
\label{para1}
\end{equation}
Squared mass of sfermions is also checked whether it satisfies the experimental
bounds. In this calculation we have to take account of 
the $D$-term contribution since it may take a large
negative value. Supersymmetry breaking parameters such as the 
soft scalar mass $m_0$ and the $A$ parameters are assumed to take a
universal value $m_{3/2}=1$~TeV.
 
At first we examine the appearance of the singlino dominated lightest
neutralino. In the panels of Fig.~1 we plot the fraction 
$|U_{\ell j}|^2$ of each 
component ${\cal N}_j$ of the lightest neutralino $\tilde\chi^0_\ell$ 
for the $Z^\prime$ mass $m_{Z_2}$, which is related to the vacuum expectation 
value $u$ through eq.~(\ref{z2}).\footnote{$m_{Z_2}$ and $u$ are 
essentially proportional to each other. 
In case of $u\gg v_{1,2}$ their proportional
constant is given as $m_{Z_2}/u\simeq g_x|Q_S|/\sqrt 2\simeq 1.26$.}
In these panels we choose two typical values of $M_x$ and 
fix $M_W$ and $\mu$ as $M_W=\mu=300$~GeV. 
These confirm that the singlino domination of the lightest neutralino 
can occur even for large values of $u$ which are required by the 
$Z^\prime$ phenomenology as long as $M_x$ is sufficiently large. 
The left panel shows that the lightest neutralino
rapidly turns from the singlino dominated one to the bino dominated one 
when $u$ reaches a certain value. 
In the right panel the lightest neutralino is dominated by the singlino
throughout the whole regions of $u$ since $M_x$ is large enough.
In these panels it is interesting that there is an upper bound for $m_{Z_2}$. 
This is caused by a condition for the mass of down type squarks since the
extra U(1) $D$-term contributions are negative for them.
It should also be noted that a value of $\mu$ is fixed in these panels. 
The coupling $\lambda$ becomes larger as we make a value of 
$u$ smaller. This explains such behavior that the Higgsino 
components $\tilde H_{1,2}$ increase in the regions of smaller $u$. 
If we make values of $M_W$ and $\mu$
larger for relatively small values of $M_x$, the regions of $m_{Z_2}$
where the lightest neutralino is dominated by the singlino
is extended upward, keeping the features shown in the left panel of Fig. 1.
   
\input epsf 
\begin{figure}[tb]
\begin{center}
\epsfxsize=7.7cm
\leavevmode
\epsfbox{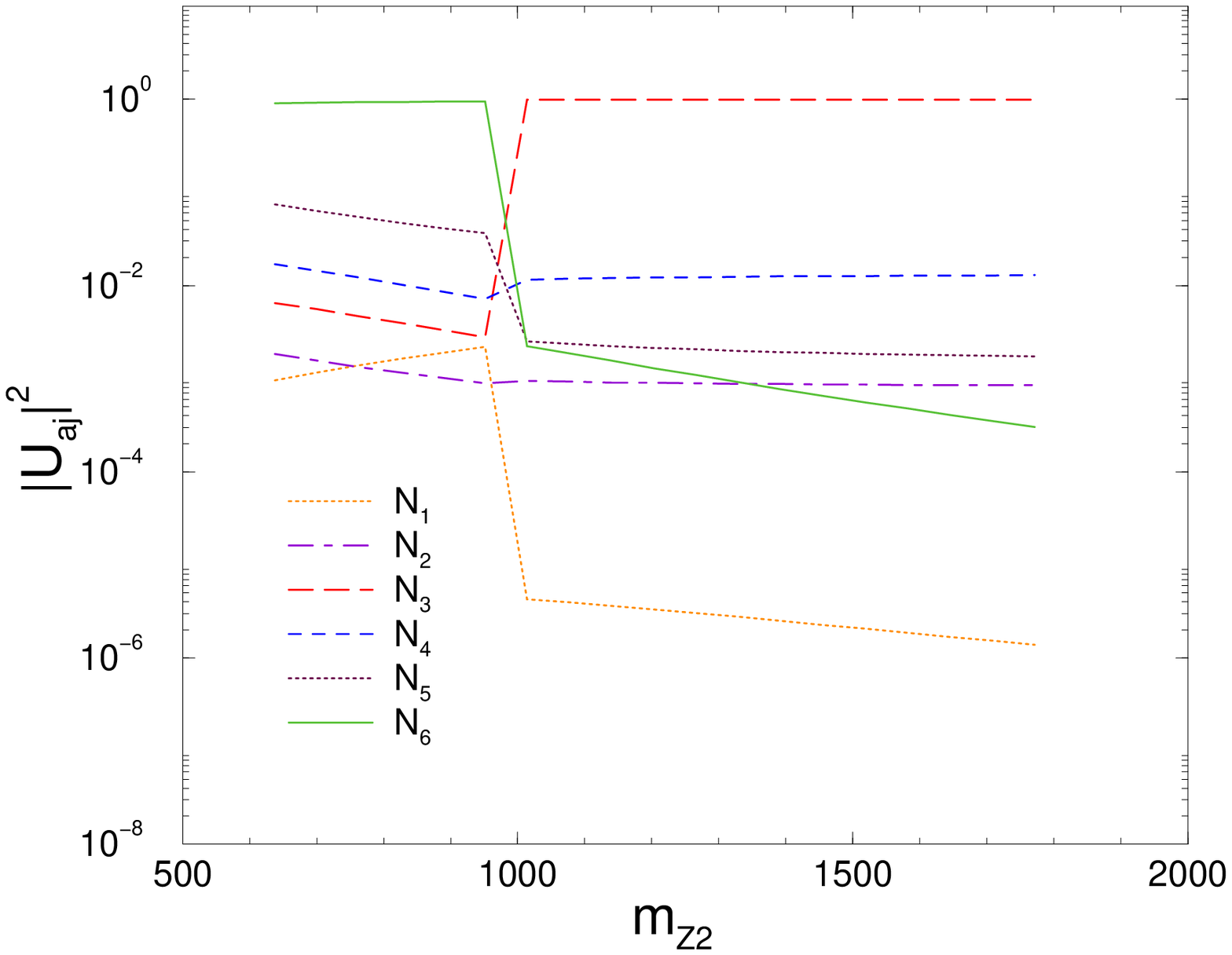}
\hspace*{3mm}
\epsfxsize=7.7cm
\leavevmode
\epsfbox{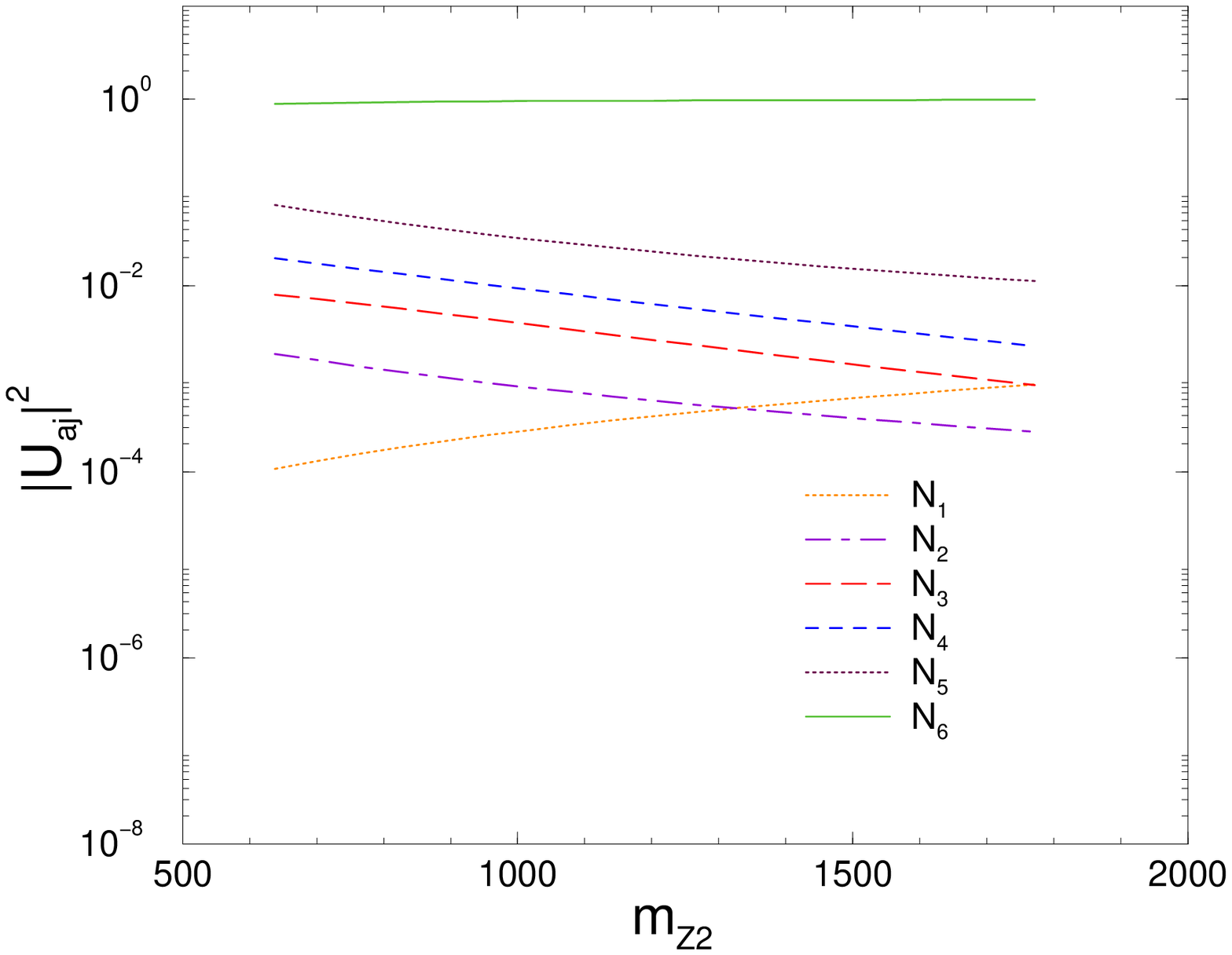}\\
\end{center}
\vspace*{-6mm}
{\footnotesize Fig. 1~~The composition $|U_{\ell j}|^2$ of the lightest
 neutralino $\tilde\chi^0_\ell$ in the case of $M_W=\mu=300$~GeV.
$M_x$ is fixed to be 20~TeV and 60~TeV in the
 left and right panel, respectively. In these panels the Higgs mass
 bound is not imposed.}
\end{figure}

In Fig.~2 we plot the mass eigenvalue of the lightest neutralino 
for $m_{Z_2}$. We choose typical values of $M_x$ and also fix 
both $M_W$ and $\mu$ to be 300~GeV and 600~GeV in the left and
right panels.
We find from these panels that the singlino dominated 
lightest neutralino can be very light.  The mass becomes 
smaller than $m_Z/2$ in certain regions of $u$ 
for sufficiently large values of $M_x$. 
Although such a light 
neutralino seems to be forbidden in the MSSM
from both the invisible $Z$ width and 
the chargino mass bound, the singlino domination makes the
model able to evade these constraints.
In both panels the small $m_{Z_2}$ regions are found to be forbidden by 
the conditions imposed on $\lambda$ and $m_{Z_2}$ in eqs.~(\ref{para})
and (\ref{para1}). 
If we remove these conditions, the right
panel will show similar behavior as those in the left panel. 
At the small $m_{Z_2}$ regions in the left panel, the mass eigenvalue 
increases as $m_{Z_2}$ takes smaller values. 
Since we fix $\mu$, the coupling $\lambda$
increases for smaller $u$ values. This makes the Higgsino components
of the lightest neutralino increase in these regions as shown in Fig. 1. 
These explain the behavior of the mass eigenvalues there. 
The left panel shows that the mass eigenvalues become
constant in the regions where $m_{Z_2}$ is larger than a certain 
value which is
determined by $M_x$. This behavior can be explained by the fact that 
the singlino domination finishes there.  
The lightest neutralino starts being dominated by the bino 
in larger $m_{Z_2}$ regions. In these regions the lightest neutralino 
has the similar nature to that in the MSSM. 
If we make $M_W$ and $\mu$ larger, these MSSM 
like regions start at a larger $m_{Z_2}$ and a corresponding 
mass eigenvalue also becomes larger as indicated in these panels.   

\begin{figure}[tb]
\begin{center}
\epsfxsize=7.7cm
\leavevmode
\epsfbox{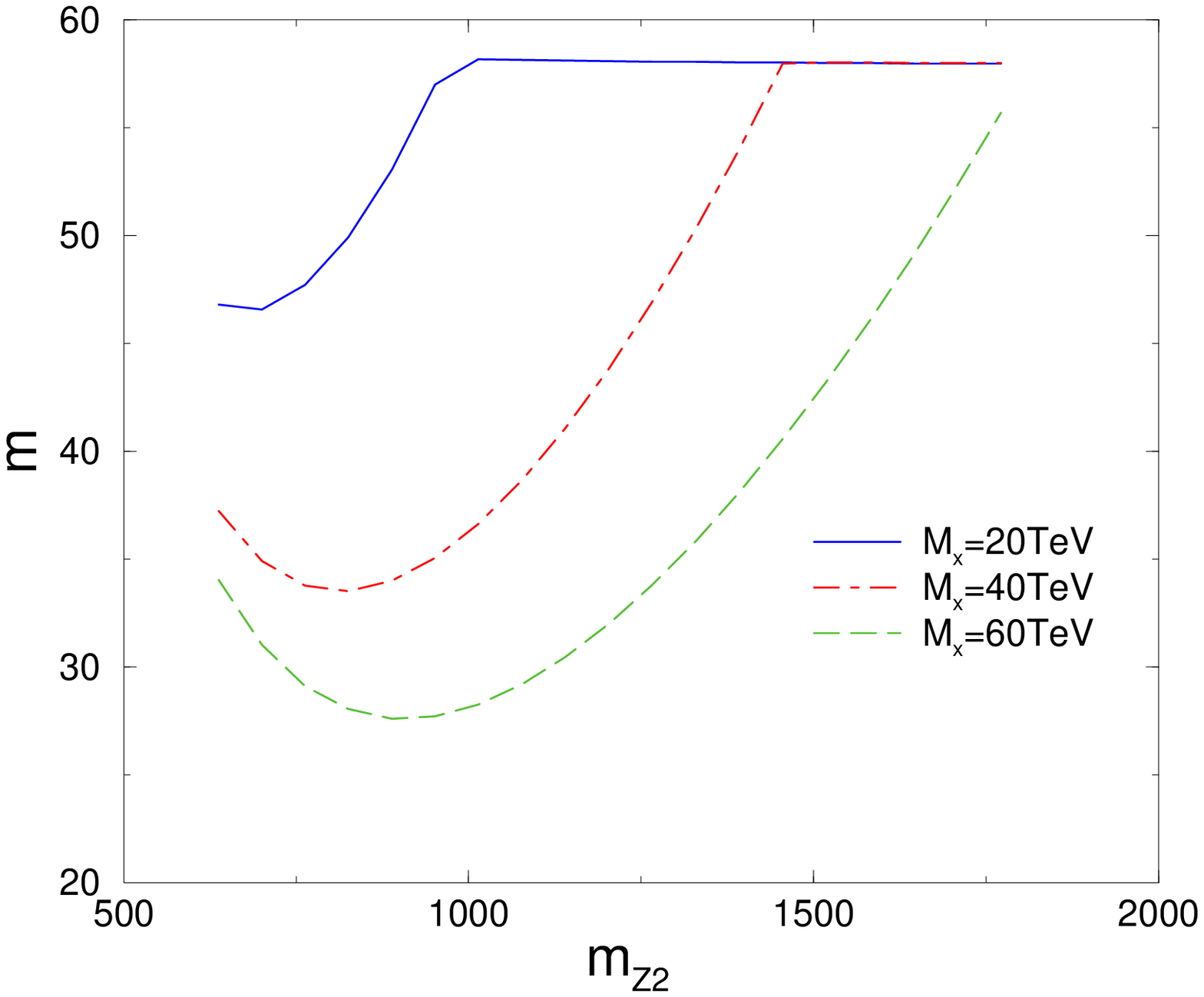}
\hspace*{3mm}
\epsfxsize=7.7cm
\leavevmode
\epsfbox{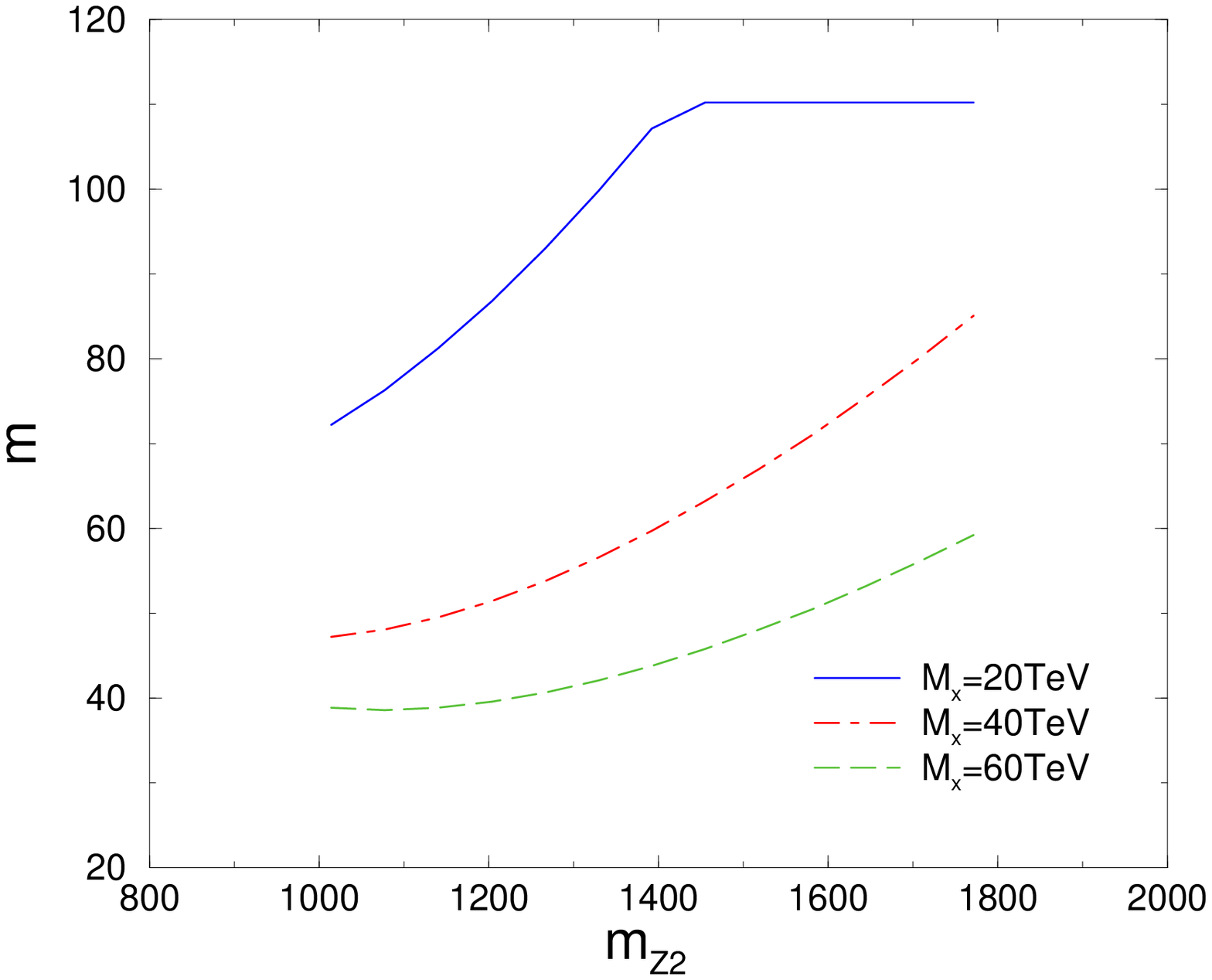}
\end{center}
\vspace*{-6mm}
{\footnotesize Fig. 2~~The mass eigenvalue $m_{\tilde\chi^0_\ell}$ 
of the lightest neutralino $\tilde\chi_\ell^0$. 
We take $M_W=\mu=300$~GeV and 600~GeV in the left 
and right panel, respectively. The Higgs mass bound is not imposed in
 these panels.}
\end{figure}

If we regard the singlino dominated neutralino as the CDM 
candidate, a very different interaction from the MSSM can 
contribute to their annihilation cross section. 
In fact, the lightest neutralino dominated by the singlino $\tilde S$ 
is expected to annihilate mainly through the $Z^\prime$ exchange. 
However, it is a new channel to be 
effective only in the case that $Z^\prime$ is not so heavy.
Although the lightest neutralino can be lighter than that in the MSSM, 
if $Z^\prime$ is much heavier than the weak scale, the singlino 
dominated lightest neutralino cannot annihilate effectively through
this mode and we may have too much relic abundance for it.
If we take account of this aspect and also the behavior of the mass
eigenvalue shown in Fig.2, we find that smaller $u$ regions seem to be favored 
for the explanation of the CDM abundance for a fixed $M_x$ .

We should also remind here that the Higgsino components and the bino
component of the lightest neutralino increase for smaller values of 
$m_{Z_2}$ in the singlino dominated LSP regions.
In these regions the annihilation may also be effectively mediated 
by the exchange of the $Z$ boson.
In fact, as found in Figs.~1 and 2, the lightest neutralino mass
$m_{\tilde\chi^0_\ell}$ can be $m_Z/2$ there for suitable parameter sets. 
Thus, the annihilation of the lightest neutralinos can be enhanced 
due to the $Z$ pole effect in the case that it has substantial 
Higgsino components.
Figs.~1 and 2 suggest that such a possibility may be realized more 
effectively in smaller $m_{Z_2}$ regions for a fixed $M_x$. 
It should be noted that this enhancement is expected to occur at the
singlino dominated LSP regions where the lightest neutralino 
does not start being similar to that of the MSSM and the NMSSM.
 
It is also useful to note that another difference of this model from 
the MSSM and the NMSSM exists in the neutral Higgs sector.
In this kind of model, the mass of the lightest neutral Higgs 
scalar has an extra U(1) $D$-term contribution compared with the 
NMSSM \cite{extra,higgsb}.
Since its upper bound can be estimated as 
\begin{equation}
m_h^2\le m_Z^2\left[\cos^22\beta+{2\lambda^2\over g_W^2+g_Y^2}\sin^22\beta+
{g_x^2\over g_W^2+g_Y^2}(Q_1\cos^2\beta+Q_2\sin^2\beta)^2\right] 
+\Delta m_1^2,
\label{higgs}
\end{equation}
it can be heavier than that of the MSSM and also the 
NMSSM. Due to the second and third terms, 
even in the regions of the small $\tan\beta$, 
the lightest neutral Higgs mass $m_h$ can take a large value 
such as 140 GeV or
more, if the one-loop correction $\Delta m_1^2$ is taken into account. 
Since its dominant components of this lightest Higgs scalar are 
considered to be $H_{1,2}^0$,
its interaction with the bino and the Higgsinos is similar to that in 
the MSSM.
Here we remind the behavior of each component $|U_{\ell j}|^2$ shown in Fig.~1
and also the mass eigenvalue $m_{\tilde\chi^0_\ell}$  
of the lightest neutralino shown in Fig.~2. 
Then we find that $m_{\tilde\chi^0_\ell}\sim m_h/2$ may 
be easily realized at a
certain value of $m_{Z_2}$ if values of $M_W$, $\mu$ and $M_x$
are chosen suitably. The lightest neutralino could have
substantial Higgsino components there although the singlino domination 
is still satisfied.   
These features suggest that the annihilation of the lightest 
neutralino mediated by the Higgs exchange may also be enhanced due to 
the Higgs pole effect. 
Although small $\tan\beta$ regions are now disfavored by the neutral
Higgs mass bound in the MSSM, such regions may still be 
interesting from a viewpoint of the CDM in the present model.
  
Finally we comment on the relation to another possibility 
of the singlino dominated lightest neutralino.
The singlino dominated lightest neutralino is known to appear in the
NMSSM and the $S$-model \cite{nmssm0,nmssm,secl}.
In these models the lightest neutral Higgs scalar may also be 
dominated by the singlet scalar since the vacuum expectation
value $u$ can take a weak scale or a smaller value. 
As the result, its mass eigenvalue can be much smaller than the 
currently known LEP2 lower bound of the neutral Higgs mass.
Since $u$ is small, the coupling $\lambda$ can take a large value. 
The Higgs exchange process can play a main role for the annihilation 
of the lightest neutralino in this case. Although the singlino dominates
the lightest neutralino also in these models, the phenomenology including the
annihilation process of the lightest neutralino is completely 
different from our models.
In our models there may be various possibilities for the mass of the 
lightest neutralino, which make it possible to consider a different type
of annihilation process.
If the CDM is found to be the singlino dominated
neutralino, we might be able to refer to its annihilation processes and
also the nature of the lightest neutral Higgs scalar to distinguish 
the present possibility from the NMSSM and the $S$-model. 

\subsection{Relic abundance of the singlino dominated neutralino}
Now we study whether the singlino dominated neutralino can be the CDM 
candidate by using numerical analyses of its relic
abundance. At first, we briefly review how to estimate the relic abundance 
of thermal plasma in the expanding universe \cite{xf,ann}. 
The relic abundance of the thermal stable lightest neutralino
$\tilde\chi^0_\ell$ can be evaluated as the thermal abundance at its
freeze-out temperature $T_F$, which can be determined by 
$H(T_F)\sim \Gamma_{\tilde\chi^0_\ell}$.
$H(T_F)$ is the Hubble parameter at $T_F$. 
$\Gamma_{\tilde\chi^0_\ell}$ is an annihilation 
rate of $\tilde\chi^0_\ell$ and it can be written as
$\Gamma_{\tilde\chi^0_\ell}=\langle\sigma_{\rm ann} v\rangle
n_{\tilde\chi^0_\ell}$, 
where $\langle\sigma_{\rm ann}v\rangle$  is thermal average of 
the product of the annihilation cross section $\sigma_{\rm ann}$
and the relative velocity $v$ of annihilating $\tilde\chi^0_\ell$s 
in the center of mass frame.
Thermal number density of non-relativistic $\tilde\chi^0_\ell$s 
at this temperature is expressed by $n_{\tilde\chi^0_\ell}$. 
If we introduce a dimensionless parameter $x_F=m_{\tilde\chi^0_\ell}/T_F$, 
we find that $x_F$ can be represented as
\begin{equation}
x_F=\ln{m_{\rm pl}m_{\tilde\chi^0_\ell}\langle\sigma_{\rm ann}v\rangle
\over 13(g_\ast x_F)^{1/2}},
\end{equation}  
where $g_\ast$ enumerates the degrees of freedom of relativistic 
particles at
$T_F$. Using this $x_F$, the present abundance of $\tilde\chi^0_\ell$  
can be estimated as
\begin{equation}
\Omega_\chi h^2|_0=
\left.{m_{\tilde\chi^0_\ell} 
n_{\tilde\chi^0_\ell}\over \rho_{\rm cr}/h^2 }\right|_0
\simeq{8.77\times 10^{-11}g_\ast^{-1/2}x_F\over 
\langle\sigma_{\rm ann}v\rangle~{\rm GeV}^2 }.
\end{equation} 
Here we may use the approximation such as 
\begin{equation}
\langle\sigma_{\rm ann}v\rangle\simeq a+(b-3a/2)/x_F
\label{annihi}
\end{equation}
under the non-relativistic expansion 
$\sigma_{\rm ann}v\simeq a+bv^2/6$.
Detailed formulas of $a$ and $b$ for annihilation processes 
induced by the exchange of various
fields can be found in \cite{ann,ann1}.

If the lightest neutralino $\tilde\chi^0_\ell$ is lighter than $m_Z$, 
only the annihilation into the SM
fermion-antifermion pairs 
$\tilde\chi^0_\ell\tilde\chi^0_\ell\rightarrow f\bar f$ is expected to occur. 
We consider this case.
The neutralino annihilation processes in the models with an extra U(1) 
are expected to be mediated by the exchange of $Z$, $Z^\prime$ 
and the neutral Higgs scalars in the $s$-channel 
and by the sfermion exchange in the $t$-channel as usual. 
However, since the singlino dominates the lightest neutralino in our
case, its annihilation cross section is expected to have 
a dominant contribution from the $Z^\prime$ exchange. 
If we define contribution of this
process to $a$ and $b$ in eq.~(\ref{annihi}) as $a_f$ and $b_f$,
they can be expressed as \cite{ce}
\begin{eqnarray}
&&\hspace*{-7mm}a_f={2c_f\over\pi}
{g_x^4m_f^2\over m_{Z_2}^4}\sqrt{1-{m_f^2\over m_{\tilde\chi^0_\ell}^2}}
\left[({Q(f_L)\over 2}-{Q(f_R)\over 2})^2
(\sum_{j=4}^6{Q_j\over 2}U_{\ell j}^2)^2\right], \nonumber \\
&&\hspace*{-7mm}b_f=\left(-{9\over 2}
+{3 \over 4}{m_f^2\over m_{\tilde\chi^0_\ell}^2-m_f^2}\right)a_f \nonumber \\ 
&&\hspace*{-2mm}+{2c_f\over\pi}  
\left({g_x^2m_{\tilde\chi^0_\ell}
\sum_{j=4}^6{Q_j\over 2}U_{\ell j}^2\over 
4m_{\tilde\chi^0_\ell}^2-m_{Z_2}^2}\right)^2 
\sqrt{1-{m_f^2\over m_{\tilde\chi^0_\ell}^2}}
\left[\left({Q(f_L)\over 2}\right)^2+ \left({Q(f_R)\over 2}\right)^2\right] 
\left(4+{2m_f^2\over m_{\tilde\chi^0_\ell}^2}\right),
\label{zprime}
\end{eqnarray} 
where $c_f=1$ for leptons and 3 for quarks. $m_{\tilde\chi^0_\ell}$ 
stands for the mass of the singlino dominated lightest 
neutralino $\tilde\chi^0_\ell$. 
The extra U(1) charges of fermions $f_{L,R}$ are denoted by $Q(f_L)$ 
and $Q(f_R)$. If we note the charge
conservation in Yukawa couplings, we find that only $Q_{1,2}$ and
$Q(f_L)$ are necessary to fix the relevant charges.
For other annihilation processes mediated by the MSSM contents, we can
find the cross section formulas in \cite{ann,ann1}.
In the numerical calculation we also take account of these.

We now show that this singlino dominated neutralino can have 
suitable relic abundance as dark matter.
In Fig.~3 we show seven regions in the $(m_{Z_2}, M_x)$ plane by surrounding
them with various kinds of lines, where the singlino dominated lightest 
neutralino is realized for fixed values of $M_W=\mu$ 
within the parameter space defined by (\ref{para}) under the 
conditions (\ref{para1}).
The value of $M_W$ is taken from 200~GeV to 800~GeV at a 100~GeV
interval.
The regions corresponding to each value of $M_W$
are plotted by a solid line (200~GeV), a dash-dotted line
(300~GeV), a dashed line (400~GeV), a dotted line (500~GeV) etc., respectively.
The vertical lines corresponding to 200 GeV and 300 GeV overlap at
$\sim$640 GeV. 
The regions appearing for a larger $m_{Z_2}$ 
correspond to the one for a larger $M_W$.   
In each region, the lower bound of $m_{Z_2}$ is determined by 
the condition for $\lambda$ given in (\ref{para}) except 
for the case of $M_W=200$~GeV, for which it comes from the $m_{Z_2}$ bound.
In all regions the upper bound of $m_{Z_2}$ takes a common value which
comes from the down type squark mass condition as mentioned before.
The lower bound of $M_x$ for each value of $m_{Z_2}$ comes from 
the requirement that the lightest neutralino is dominated by the singlino. 

\begin{figure}[tb]
\begin{center}
\epsfxsize=7.7cm
\leavevmode
\epsfbox{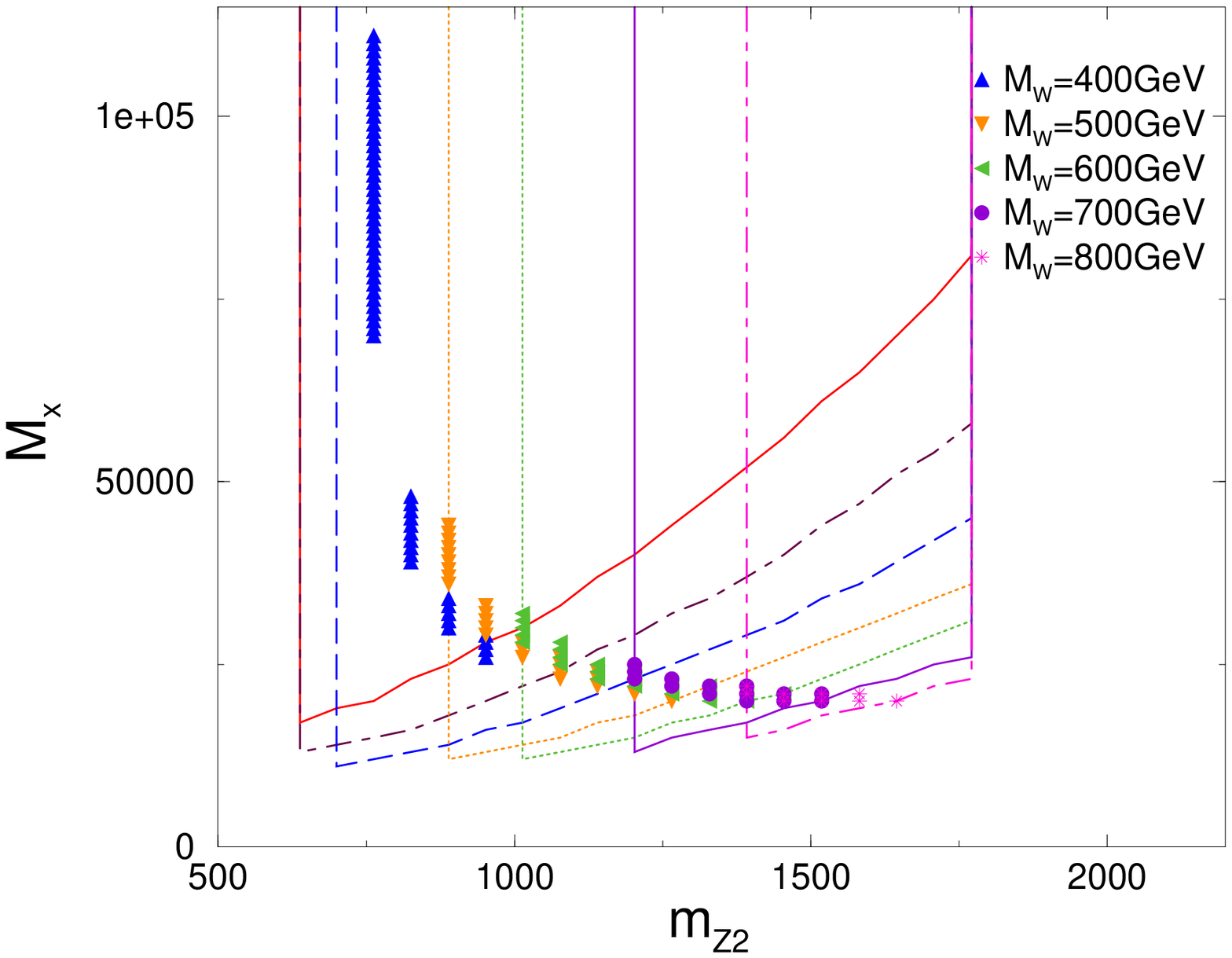}
\hspace*{3mm}
\epsfxsize=7.7cm
\leavevmode
\epsfbox{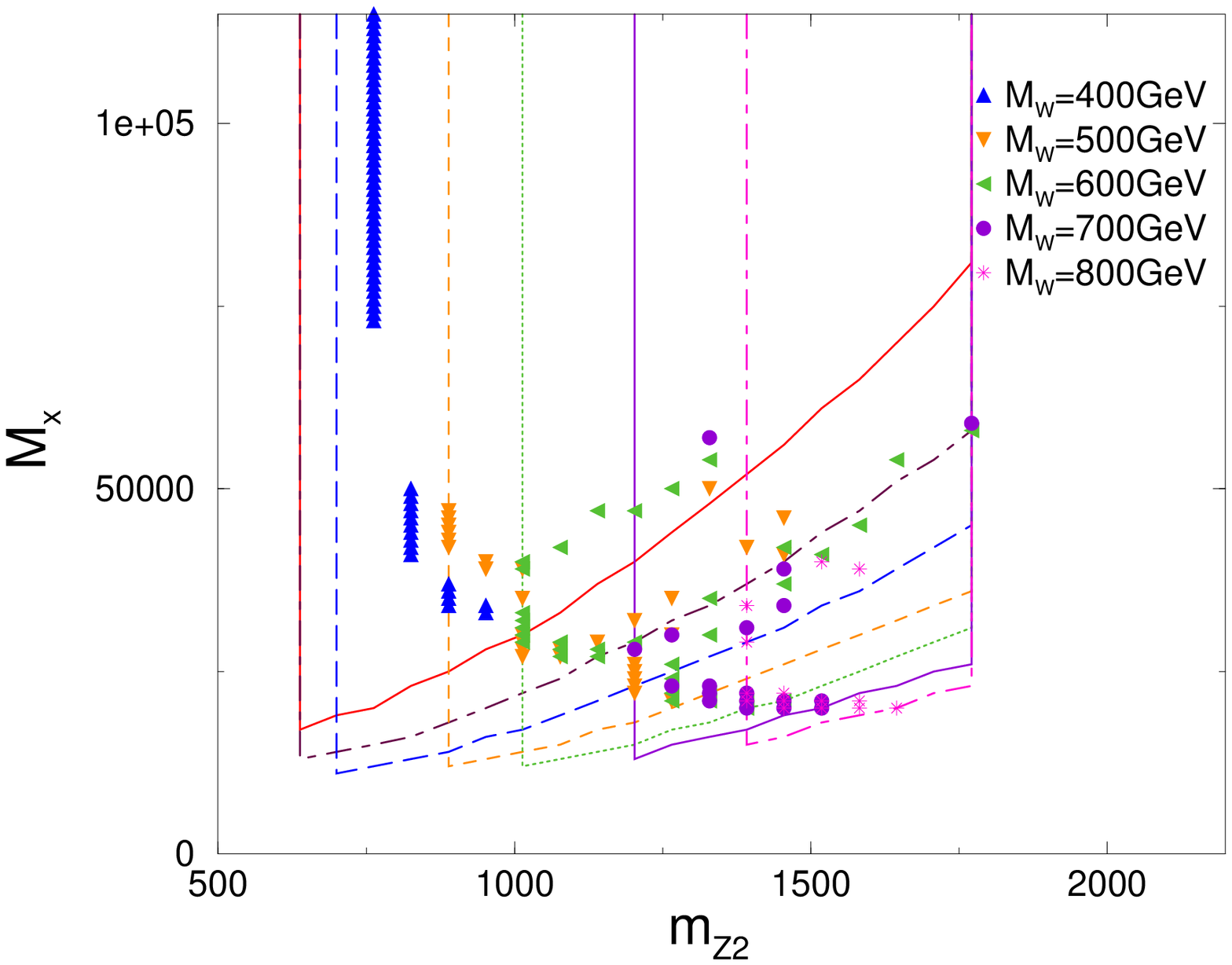}\\
\end{center}
\vspace*{-6mm}
{\footnotesize Fig. 3~~The regions in the $(m_{Z_2}, M_x)$ plane where 
the singlino dominated lightest neutralino is realized. 
Points which satisfy the CDM constraints from 
the WMAP are plotted for various values of $M_W=\mu$. 
Only the $Z^\prime$ exchange is taken into account in the left panel. 
In the right panel all processes in the MSSM are also included.}
\end{figure}

In the left panel of Fig. 3  we also plot points 
where the $\Omega_{\rm CDM}$ constraint from the WMAP is satisfied 
for various values of $M_W=\mu$. In this panel we only take account of
the $Z^\prime$ exchange process described by eq.~(\ref{zprime}). 
These points are found by scanning $u$ and $M_x$ at the interval of
50 GeV and 1000 GeV, respectively. 
These solutions are obtained for $x_F\simeq 22-23$.
The $\Omega_{\rm CDM}$ solutions for $M_W=200$~GeV and 300~GeV are
excluded by the condition for the Higgs mass $m_h\ge 114$~GeV. 
This panel shows that each point satisfying the WMAP constraint 
is found only in the singlino dominated LSP regions.
This seems consistent with the discussion given in 
the previous part. 

\begin{figure}[htb]
\begin{center}
\begin{tabular}{c|ccccc}\hline
$M_W$ & 400  & 500 & 600 & 700 & 800 \\\hline\hline
$m_{\tilde\chi^0_\ell}$ & 31--44 & 40--87 & 45--96 & 45--120& 70--144 \\  
$m_h$ & 116--129 & 115--146 & 117--154 & 127--154 & 142--153 \\
\end{tabular}
\end{center}
\vspace*{2mm}
{\footnotesize Table~1~~$m_{\tilde\chi^0_\ell}$ 
and $m_h$ which give the solutions in the right panel of Fig.~3 
for the CDM constraints from the WMAP. The mass unit is GeV.} 
\end{figure}

In the right panel we plot points where the $\Omega_{\rm CDM}$
constraint from the WMAP is satisfied by including all the annihilation
processes in the MSSM such as the $Z$ boson exchange, the Higgs scalar 
exchange and so on.\footnote{In this calculation, as an example, we take 
$\sqrt{\Delta m_1^2}$ as 80~GeV. Even if we change this
value,  qualitatively similar results can be obtained.}
These solutions are obtained for $x_F\simeq 22-23$. For these solutions 
both the mass of the lightest neutralino $m_{\tilde\chi^0_\ell}$ and 
the lightest Higgs $m_h$ are shown in Table~1.
The Higgs mass decreases for larger values of $u$.
This is expected from eq.~(\ref{higgs}), since the second term 
in the brackets of the right-hand side of eq.~(\ref{higgs}) 
decreases as increasing $u$ for a fixed $\mu$.  
This feature forbids the solutions in larger $u$ regions in the case of 
smaller values of $M_W$. 
Table~1 shows that there are possibilities such that $m_{\tilde\chi^0_\ell}$
takes values near $m_Z/2$ or $m_h/2$ in the case of $M_W\ge500$~GeV.
Corresponding to these, we can find that additional solutions appear in
this panel compared with the left panel.
These solutions may be understood along the line discussed already.
They are considered to appear as the result of the additional effect
caused by the enhancement of the annihilation due to the $Z$ pole or the 
lightest neutral Higgs pole. 
Also in this panel, we find that all solutions appear only in the regions
which satisfy the singlino domination condition.
We can also find these qualitative features for other parameter sets.

From these figures we find that the singlino dominated lightest neutralino 
can be a good CDM candidate. Its annihilation can be mediated 
by various processes. This is considered to be caused by 
the feature in the present model that the singlino dominated 
lightest neutralino can have its mass eigenvalue in a rather wide range.

In the above study we impose $m_{Z_2}\ge 600$~GeV on the $Z^\prime$
mass for simplicity, although this bound depends on the models.
We need to confirm that our solutions are consistent with the 
result of the direct search of $Z^\prime$ at the Tevatron. 
The CDF limit at $\sqrt{s}=1.8$~TeV is expressed as \cite{cdf} 
\begin{equation}
\sigma(p\bar p\rightarrow Z^\prime X)B(Z^\prime\rightarrow 
e^+e^-,~\mu^+\mu^-)<0.04~{\rm pb}.
\end{equation}
We calculated this $\sigma B$ for the dilepton modes for each solution 
shown in the right panel of Fig. 3.
Then we found that all solutions satisfied this CDF bound. 
In the present model $Z^\prime$ can have a rather large branching 
ratio such as 15-19\% into the neutralino sector. 
   
We comment on a desired feature for models in order for
the present scenario to work well finally. As mentioned already,
the strength of the extra U(1) interaction is an important factor for the
annihilation of the singlino dominated neutralino caused by the
$s$-channel exchange of $Z^\prime$. 
It is related to the normalization constant 
$k$ or the extra U(1) charge of various fields. Since $k=O(1)$ is naturally
expected, the charge $|Q_S|$ of the singlet chiral superfield $\hat S$
should be larger compared with those of other fields.
If we assume to satisfy $\tan\beta\simeq\sqrt{Q_1/Q_2}$ so as   
to evade the constraint from the $ZZ^\prime$ mixing independently of the
$Z^\prime$ mass, the charge conservation 
automatically makes $|Q_S|$ larger than the charges 
$|Q_{1,2}|$ of the Higgs doublets $H_{1,2}$. 
Thus, as stressed before, the annihilation of the singlino
dominated lightest neutralinos is expected to be enhanced as its singlino
component increases. 
If these features are satisfied, the present scenario 
seems to favor a small $\tan\beta$ generally. 

A large value of $|Q_S|$ makes the coupling of the
singlino dominated lightest neutralino with $Z^\prime$ larger.
This means that the branching ratio of the $Z^\prime$ decay may also be 
dominated by the decay modes into the singlino dominated neutralinos. 
If this happens, the current mass bound of $Z^\prime$ may be 
relaxed and allowed parameter regions can be extended. 
Although these aspects are strongly dependent on models, 
it seems worthy of proceeding with the detailed studies in more realistic
cases derived from the $E_6$ model and examining the possibility to find
$Z^\prime$ at the LHC.
It may also be interesting to study whether we can construct this 
kind of models on the basis of a fundamental framework like string theory. 

\section{Summary}
We studied a possibility that the singlino dominates the lightest
neutralino in the supersymmetric models with an extra U(1), which give 
an elegant weak scale solution for the $\mu$ problem. For that purpose 
we assumed a supersymmetry breaking scenario which induces non-universal
mass for the extra U(1) gaugino. 
When the extra U(1) gaugino is very heavy as compared with the 
other gauginos, whose mass is kept in the ordinary range,
the lightest neutralino can be shown to be dominated by 
the singlino even if the vacuum expectation value $u$ of the singlet scalar
field is large enough to permit the extra U(1) gauge field $Z^\prime$ 
to satisfy the experimental constraints.
This possibility is very different from what happens in 
the usual models with an extra U(1), where 
the lightest neutralino is expected to have the similar nature 
to that of the MSSM because of the large $u$.
In the NMSSM and a special type of models with an extra U(1),
the singlino dominated lightest neutralino is known 
to appear in the case of a small vacuum expectation value $u$ 
unless the coupling $\lambda$ is very small. 
However, our model can realize the singlino dominated
neutralino in a very different manner from those. 

We also studied whether this singlino dominated 
lightest neutralino can have the suitable relic density as 
the CDM candidate based on the WMAP results.
We found that it can be a nice CDM candidate in non-negligible 
parameter regions as long as the model satisfies
$\tan\beta\simeq\sqrt{Q_1/Q_2}$, which comes from the $ZZ^\prime$ 
mixing constraint.
This CDM candidate has very different nature from that in the MSSM 
and the NMSSM. The model might be distinguished through the studies
of phenomena related to the neutralino and the neutral Higgs scalar.
Detailed studies of these aspects seem very interesting. 
We will present such studies in a different publication. 

\vspace*{5mm}
The author would like to thank G.~Zoupanos for reading the manuscript
and giving useful comments.
This work is partially supported by a Grant-in-Aid for Scientific
Research (C) from the Japan Society for Promotion of Science (No.17540246).

\vspace*{10mm}
\noindent
{\Large\bf Appendix}

In usual supersymmetry breaking scenarios, gaugino mass 
is universal. If we consider much larger gaugino mass compared 
with the weak scale in a universal gaugino mass framework, 
the mass of the gluinos and the winos also becomes large to defeat 
the SM gauge coupling unification, for example. The gaugino mass 
universality also imposes severe constraints on phenomenological 
features of the model. If we assume the universality of
the gaugino mass and also the coupling unification, the mass of the 
gauginos in the MSSM satisfies the unification relation such as
$M_g/g_s^2=M_W/g_W^2=5M_Y/3g_Y^2$. Since the current 
lower bound of the chargino mass is shown to be 104~GeV \cite{chargino}, 
$M_W$ is difficult to be smaller than 100~GeV. This fact together 
with the unification relation constrains the allowed regions of $M_Y$. 
Thus, the lightest neutralino cannot be so light under these requirements 
in the MSSM. In the models with an extra U(1) which are not 
a type of the $S$-model, 
the situation is similar to this as long as the gaugino 
mass is assumed to be universal.

A few examples which can realize non-universal gaugino mass have been 
proposed by now.\footnote{ 
The mass of the gauginos is known to be non-universal in some kinds of
models, for example, 
in the multi-moduli supersymmetry breaking \cite{multim}, the intersecting 
D-brane models \cite{dbrane} and a certain type of gauge mediation model
\cite{s}. Phenomenological effects of the
non-universal gaugino mass on the neutralino sector is also 
studied in \cite{cm} in a different context from ours.} 
In those cases, however, non-universality is not so large that it seems to 
be difficult to make the lightest neutralino be dominated by the singlino
unlike the models assumed in the text.
Here we propose a new scenario which makes an Abelian gaugino mass 
largely different from others and then the singlino dominated neutralino 
the lightest one. 

It is known that kinetic term mixing can generally appear 
among the Abelian gauge fields in multi U(1)s 
models \cite{mixing,oneloop,mixing1}.
In the following, such mixing is assumed to exist between two Abelian 
gauge fields, each of which belongs to the hidden and observable sector.
In that case we show that there can be an additional contribution to 
the corresponding Abelian gaugino mass in the observable sector, 
if we make some assumptions on the superpotential and also 
the supersymmetry breaking in the hidden sector.   
This additional contribution may make the Abelian
gaugino mass different from others in the observable sector.
 
For simplicity, we consider a supersymmetric U(1)$_a\times$U(1)$_b$ model
where U(1)$_a$ and U(1)$_b$ belong to the hidden sector and the observable 
sector, respectively.
We suppose that $\hat W_{a,b}^\alpha$ is a chiral superfield with a spinor
index $\alpha$, which contains a field strength of U(1)$_{a,b}$. 
Since $\hat W_{a,b}^\alpha$ is gauge invariant, 
gauge invariant kinetic terms can be expressed as
\begin{equation}
{\cal L}_{\rm kin}=\int d^2\theta\left({1\over 32} 
\hat W_a^\alpha\hat W_{a\alpha} 
+{1\over 32}\hat W_b^\alpha\hat W_{b\alpha} 
+ {\sin\chi\over 16}\hat W_a^\alpha\hat W_{b\alpha} \right),
\label{kinetmix}
\end{equation}
where it should be reminded that a mixing term is generally allowed 
at least from a viewpoint of symmetry.
Although some origins such as string one-loop effects may be considered for 
this mixing term \cite{oneloop}, we do not go further into this
issue here but we only treat $\sin\chi$ in eq.~(\ref{kinetmix}) 
as a free parameter.

This mixing can be resolved by practicing the transformation
\cite{mixing,kmix1}
\begin{equation}
\left(\begin{array}{c}\hat W_a^\alpha \\ \hat W_b^\alpha \end{array}\right)
=\left(\begin{array}{cc} 1 & -\tan\chi \\ 0 & 1/\cos\chi \\\end{array}
\right)\left(\begin{array}{c}\hat W_h^\alpha \\
\hat W_x^\alpha \end{array}\right).
\label{base}
\end{equation}
If we use a new basis $(\hat W_h^\alpha, \hat W_x^\alpha)$, 
the covariant derivative
in the observable sector can be written as
\begin{equation}
D^\mu=\partial^\mu+i\left(-g_aQ_a\tan\chi+{g_bQ_b\over\cos\chi}
\right)A_x^\mu.
\end{equation}
This shows that the gauge field $A_x^\mu$ in the observable sector can
interact with the fields having a nonzero charge $Q_a$
in the hidden sector.
However, since such fields are generally considered to be heavy
enough and $\sin\chi$ is expected to be small, we can safely expect 
that there is no phenomenological contradiction at the present stage. 

Here we consider that the Abelian gauginos in both sectors obtain mass 
through the supersymmetry breaking in the hidden sector such as 
\begin{equation}
{\cal L}_{\rm gaugino}^m=m_a\tilde\lambda_a\tilde\lambda_a
+m_b\tilde\lambda_b\tilde\lambda_b,
\label{mgauge}
\end{equation}
where the mass $m_b$ of the gaugino in the observable sector may 
be considered as the ordinary universal mass $m_{1/2}$. 
If we can assume that $m_a \gg m_b$ is satisfied,
these mass terms are rewritten by using the new basis (\ref{base}) as follows,
\begin{equation}
\tilde{\cal L}_{\rm gaugino}^{m} 
=m_a\tilde\lambda_h\tilde\lambda_h+(m_b
+m_a\sin^2\chi)\tilde\lambda_x\tilde\lambda_x,
\label{modmass}
\end{equation}
where we also use $\sin\chi\ll 1$ in this derivation.
This suggests that the Abelian gaugino mass in the observable sector 
can have an additional 
contribution due to the Abelian gauge kinetic term mixing with the
gaugino in the hidden sector.
This new contribution can be a dominant one when 
the supersymmetry breaking in the hidden sector satisfies $m_a\sin^2\chi>m_b$. 
In this case the universality of the mass of gauginos in the observable
sector can be violated in the Abelian part.

We present an example for the supersymmetry breaking scenario 
which can satisfy the above mentioned condition in a framework of the 
gravity mediation supersymmetry breaking. 
We consider a hidden sector which contains the chiral superfields 
$\hat\Phi_{1,2}$ having a nonzero charge of U(1)$_a$. 
It is also supposed that the model contains 
various neutral chiral superfields like a modulus, which are represented by
$\hat M$ together. They are defined as dimensionless fields. 
Matter superfields in the observable sector are denoted by $\hat\Psi_I$.
The K\"ahler potential and the superpotential relevant to the
present argument are supposed to be written as\footnote{For simplicity,
we assume minimal kinetic terms for the matter fields.}
\begin{eqnarray}
&&{\cal K}=\kappa^{-2}\hat K(\hat M)+\hat\Phi_1^\ast\hat\Phi_1
+\hat\Phi_2^\ast\hat\Phi_2+\hat\Psi_I^\ast\hat\Psi_I+\dots, \nonumber\\
&&{\cal W}=\hat W_0(\hat M)+\hat W_1(\hat M)\hat\Phi_1\hat\Phi_2
+\hat Y_{IJK}(\hat M)\hat\Psi_I\hat\Psi_J\hat\Psi_K+\dots,
\end{eqnarray} 
where $\kappa^{-1}$ is the reduced Planck mass and 
$Q_a(\hat\Phi_1)+Q_a(\hat\Phi_2)=0$ is assumed. 
As a source relevant to the supersymmetry breaking in the hidden 
sector, we adopt a usual assumption in the case of the gravity mediation 
supersymmetry breaking. That is, the supersymmetry breaking effect is
assumed to be parameterized by \cite{sgrasusy}
\begin{equation}
F_M\equiv \kappa^2 e^{K/2}\left(W_0\partial_M K
+ \partial_M W_0\right),
\label{susyb}
\end{equation}
which is supposed to be $O(m_{3/2})$ as long as the vacuum energy 
is assumed to vanish.
The gravitino mass $m_{3/2}$ is defined by $m_{3/2}\equiv\kappa^2e^{K/2}W_0$.

Applying this assumption to the scalar potential formula in the
supergravity, we can have well known soft supersymmetry breaking 
terms of $O(m_{3/2})$ in the observable sector \cite{sgrasusy}.
The gaugino mass is generated as \cite{sgrag}
\begin{equation}
m_{1/2}={1\over 2Re[f_A(M)]}F^M\partial_M f_A(M),
\end{equation}
where $f_A(M)$ is a gauge kinetic function for the gauge 
factor group $G_A$. If $f_A(M)$ takes the same form for each factor group,
 universal gaugino mass is generated and takes a value of
$O(m_{3/2})$. This is the ordinary scenario.
In the present case, the gaugino mass $m_b$ in eq.~(\ref{mgauge}) 
is also expected to be induced by this gravity mediation and take 
the universal value $m_{1/2}$.

On the other hand, the gaugino mass $m_a$ in the hidden sector 
is generated by the mediation of the charged chiral 
superfields $\hat\Phi_{1,2}$ due to 
the second term in ${\cal W}$ as in the gauge mediation supersymmetry 
breaking scenario \cite{gmed}. 
Since it can be generated by one-loop diagrams which have the
component fields of $\hat\Phi_{1,2}$ in internal lines, it is
approximately expressed as
\begin{equation}
m_a={g_a^2\over 16\pi^2}{\langle F_1\rangle \over\langle S_1\rangle},
\label{mgauge1}
\end{equation}
where we define that $S_1$ and $F_1$ are the scalar and auxiliary
component of $\hat W_1$, respectively.
Since we are considering the gravity mediation supersymmetry breaking,
a supersymmetry breaking scale in the hidden sector should be large 
as expected from eq.~(\ref{susyb}).
It may be natural to assume that 
$\langle F_1\rangle=O(\kappa^{-1}m_{3/2})$ and
$\langle S_1\rangle=O((\kappa^{-1}m_{3/2})^{1/2})$.
If we use these values in eq.~(\ref{mgauge1}), we find that 
the gaugino $\tilde\lambda_h$ in the hidden sector obtains the mass
\begin{equation} 
m_a={g_a^2\over 16\pi^2}O((\kappa^{-1}m_{3/2})^{1/2}).
\end{equation}
Since this $m_a$ can be much larger than the ordinary gravity mediated
contribution $m_b$, the additional contribution $m_a\sin^2\chi$ 
to the Abelian gaugino mass in eq.~(\ref{modmass}) 
can break the gaugino mass universality in the observable sector. 
In fact, since $\sin\chi$ has a suitable value 
such as $\chi=O(10^{-1})$,\footnote{The 
string one-loop effects may bring this order of mixing as 
discussed in \cite{oneloop}.}  
we can expect that $m_a\sin^2\chi > m_b$ is realized 
and the Abelian gaugino mass characterized by $m_a\sin^2\chi$
can take a much larger value than other universal ones 
$O(m_{1/2})$.\footnote{We should note that an opposite case might
also be possible. In fact, if the absolute values of $m_b$ and
$m_a\sin^2\chi$ are the same order, two contributions
may substantially cancel each other to realize much smaller 
value than $m_{1/2}$.}


\end{document}